# Self-Assembled hBN Wrinkles as Planar Optofluidic Channels

*Xiliang Yang,*[†] *Tetsuo Martynowicz,*[†] *Allard Katan,*[‡] *Kenji Watanabe,*[§] *Takashi Taniguchi,*[§] *Sabina Caneva*[†]*

† Department of Precision and Microsystems Engineering, Delft University of Technology, Mekelweg 2, 2628 CD, Delft, The Netherlands.

‡ Kavli Institute of Nanoscience Delft, 2628 CJ Delft, The Netherlands

§ National Institute for Materials Science, 1-1 Namiki, Tsukuba, Ibaraki 305-0044 Japan

* Corresponding author. Email: s.caneva@tudelft.nl

ABSTRACT

Optically accessible, scalable planar nanofluidic channels are attractive for studying transport and localization under confinement. Two dimensional (2D) materials provide large area, atomically flat interfaces for generating such platforms, yet achieving long range one-dimensional (1D) confinement with top-down nanofabrication remains challenging because it requires reproducible nanoscale feature control over extended distances, high yield, and low nonspecific adsorption of analytes under aqueous conditions. Here we demonstrate that thermally induced wrinkling of exfoliated hexagonal boron nitride (hBN) produces self-assembled, liquid-accessible, channel-like networks through a lithography-free process. By varying flake thickness and substrate choice, we quantify statistical trends in wrinkle density and morphology, thereby establishing a practical fabrication design space. Atomic force microscopy and electron microscopy reveal wrinkle-derived geometries with vertical confinement ranging from $<2$ nm to $>100$ nm depending on flake thickness and substrate. We further employ time-sequence optical imaging upon droplet-contact, which together with Raman mapping of the water OH-stretch band and capacitance-gradient mapping (dC/dz) by scanning dielectric microscopy (KPFM-based) measurements, demonstrates liquid infiltration and long-term liquid retention within the wrinkle network for more than 10 h. We finally show a proof-of-concept biomolecule confinement application in which we integrate a graphene overlayer as a background suppression interface, enabling wide-field fluorescence localization of ATTO647N labeled DNA along hBN wrinkle-induced nanochannels. Overall, this work establishes self-assembled hBN wrinkles as a scalable, and optically addressable planar nanofluidic platform for confinement of fluids and biomolecules.

## Introduction

Microfluidic systems enable robust manipulation of small liquid volumes, supporting controlled delivery, mixing, and long-term observation of chemical and biological processes under well defined boundary conditions [1]. As the field has matured, increasing attention has shifted toward nanofluidic regimes, where transport is governed not only by bulk hydrodynamics but also by interfacial effects such as surface slip, electrostatic interactions, adsorption desorption kinetics, and confinement dependent mobility [2, 3]. Developing platforms that bridge practical microfluidic handling with planar nanoscale confinement remains a central goal for lab on a chip engineering and molecular diagnostics [4].

A persistent challenge is that conventional top-down nanofabrication, while capable of producing geometrically well-defined nanochannels, often requires a trade-off in terms of yield, surface quality, and integration complexity [4]. Etching and high energy patterning can introduce roughness, chemical residues, and charge traps that complicate quantitative transport measurements and aggravate nonspecific adsorption in aqueous environments [3, 4]. These limitations become increasingly severe when long, high aspect ratio channels are required or when the device must remain compatible with optical readout and surface sensitive characterization.

2D materials offer an attractive alternative route to planar nanoconfinement because they provide atomically flat, chemically stable interfaces that can form exceptionally smooth channel walls [5]. A particularly effective strategy is van der Waals assembly of layered crystals to create slit like capillaries whose height is set by the thickness of a spacer crystal with atomic layer precision [6]. Such 2D capillaries have enabled reproducible studies of molecular transport under extreme confinement and revealed behaviors that are difficult to access with conventional fabricated channels [6, 7]. Atomically flat walls can lead to unusually low friction and even ballistic like gas transport under appropriate conditions, highlighting how boundary scattering and surface specularity dominate at these length scales [7]. Van der Waals assembly can also generate very large interfacial pressures in trapped nanoscopic volumes, reported to reach the gigapascal range, underscoring that confinement can create unusual thermodynamic and physicochemical conditions inside 2D cavities [8]. Recent microfluidic compatible device architectures typically interface these confined slits to lithographically defined access regions and reservoirs, enabling controlled filling and quantitative friction and transport measurements [9]. At the same time, there is strong motivation to reduce contamination and processing overhead in these platforms, for example by using stencil-mask (resist-free) fabrication routes that enable rapid assembly of clean van der Waals nanochannels[10].

An underexplored and potentially scalable alternative is to repurpose wrinkles, which are ubiquitous in 2D material processing, as functional channel like confinements. Wrinkling occurs when a thin film or 2D crystal on a substrate experiences compressive stress and relaxes by buckling to reduce its strain energy, a phenomenon broadly studied across soft matter and thin film mechanics [11]. In 2D materials, wrinkling and folding have been explored not only as processing outcomes but also as tunable morphological degrees of freedom that can be engineered through substrate interactions and thermal or mechanical loading [12]. Importantly, wrinkles need not be purely random disorder. In few layer hexagonal boron nitride, thermal annealing was shown to generate crystal oriented wrinkle networks with predominantly threefold and occasionally fourfold origami type junctions, indicating that wrinkling can yield structured and crystallographically registered pathways [13]. From a microfluidics viewpoint, wrinkle defined pathways are appealing because they can be formed over large areas without aggressive etching, remain intrinsically planar and optically accessible, and naturally generate extended networks that can be interrogated by microscopy and scanning probe methods.

Among van der Waals materials, hexagonal boron nitride (hBN) is a compelling candidate for wrinkle-based planar nanoconfinements. It is chemically robust, optically transparent, and widely used as an atomically flat support and encapsulation layer in 2D heterostructures [14]. Its anisotropic thermal expansion, including negative in plane thermal expansion over relevant temperature ranges, makes thermal expansion mismatch with common substrates a practical route to generate compressive stress and wrinkling during annealing and cooling [15]. Importantly, these wrinkles do not only define a mechanical morphology, but also create planar nanoslits that can host water/electrolytes under extreme confinement. Once liquids enter such nanoslits, the confined state itself becomes a central question, because nanoconfined water and electrolytes can exhibit dielectric and transport behavior that deviates strongly from bulk due to interfacial polarization, charge screening, and confinement-dependent ion mobility [15]. This motivates direct identification and characterization of the liquid-filled state in wrinkle-defined nanoconfinements as a key objective for planar nanofluidic platforms.

In this work, we demonstrate that thermally induced wrinkling of exfoliated hBN flakes produces self-assembled and liquid-accessible channel networks. By varying substrate choice and flake thickness, we map how thermal expansion mismatch and bending rigidity shape wrinkle density and channel morphology, thereby defining a nanolithography-free practical design space. We correlate structural characterization with optical readout and scanning probe measurements to assess liquid accessibility and long-term retention in the wrinkle network. Finally, we show how an added graphene overlayer can serve as an optical background suppression interface for wide-field fluorescence imaging, leveraging the strong distance dependence of nonradiative energy transfer from emitters to graphene [16, 17]. Fluorescently labeled biomolecules are used as proof-of-concept tracers to illustrate localization along wrinkle-defined pathways, thereby also addressing applications in analyte transport dynamics and in situ metrology.

## Results and discussion

### Wrinkle Network Formation and Morphology

Figure 1a summarizes the workflow used to generate wrinkle-defined channels in multilayer hBN. Multilayer hBN flakes are mechanically exfoliated onto $SiO_2$/Si, sapphire, or quartz substrates and then annealed at 1000 °C in high vacuum (~$7.4 \times 10^{-5}$ mbar). During cooling, the thermal expansion mismatch between hBN and the supporting substrate produces an effective in plane compressive strain in the flake, which is released by out of plane buckling, yielding isolated ridges as well as interconnected networks (Figures 1b and 1c). Notably, the resulting networks are not purely random. Wrinkle segments are frequently straight over micron length scales and form junctions with angles close to 120° (Figure 1d), consistent with crystal oriented wrinkle networks and origami type junctions previously reported for few layer hBN after annealing [13, 18]. Prior structural studies indicate that wrinkle orientation in multilayer hBN are often crystallographically registered and commonly align with the armchair direction [18]. This supports the view that crystallography and anisotropic bending energetics bias the pathway directions during stress relaxation.

To quantify the accessible parameter space, we measured wrinkle density and geometry across substrates and thickness ranges. Wrinkle density, defined as the total traced wrinkle length per unit area, decreases systematically from sapphire to $SiO_2$/Si to quartz (Figure 1e), in line with the expectation that a smaller effective thermal expansion mismatch reduces the driving force for wrinkle nucleation [19]. The TEC of hBN and the substrates used are reported in Table S1 in the Supporting Information. In addition, within each substrate family, thicker flakes show lower wrinkle density (Figure 1e). This trend is consistent with increased bending rigidity and stronger interlayer coupling in thicker multilayer stacks, which penalize out of plane deformation and can suppress wrinkle formation for a given applied compressive strain [20]. More broadly, the interplay between bending energy and van der Waals interactions is known to generate thickness dependent wrinkling regimes in 2D materials, reinforcing the idea that thickness acts as a practical tuning knob rather than a purely geometric parameter [20].

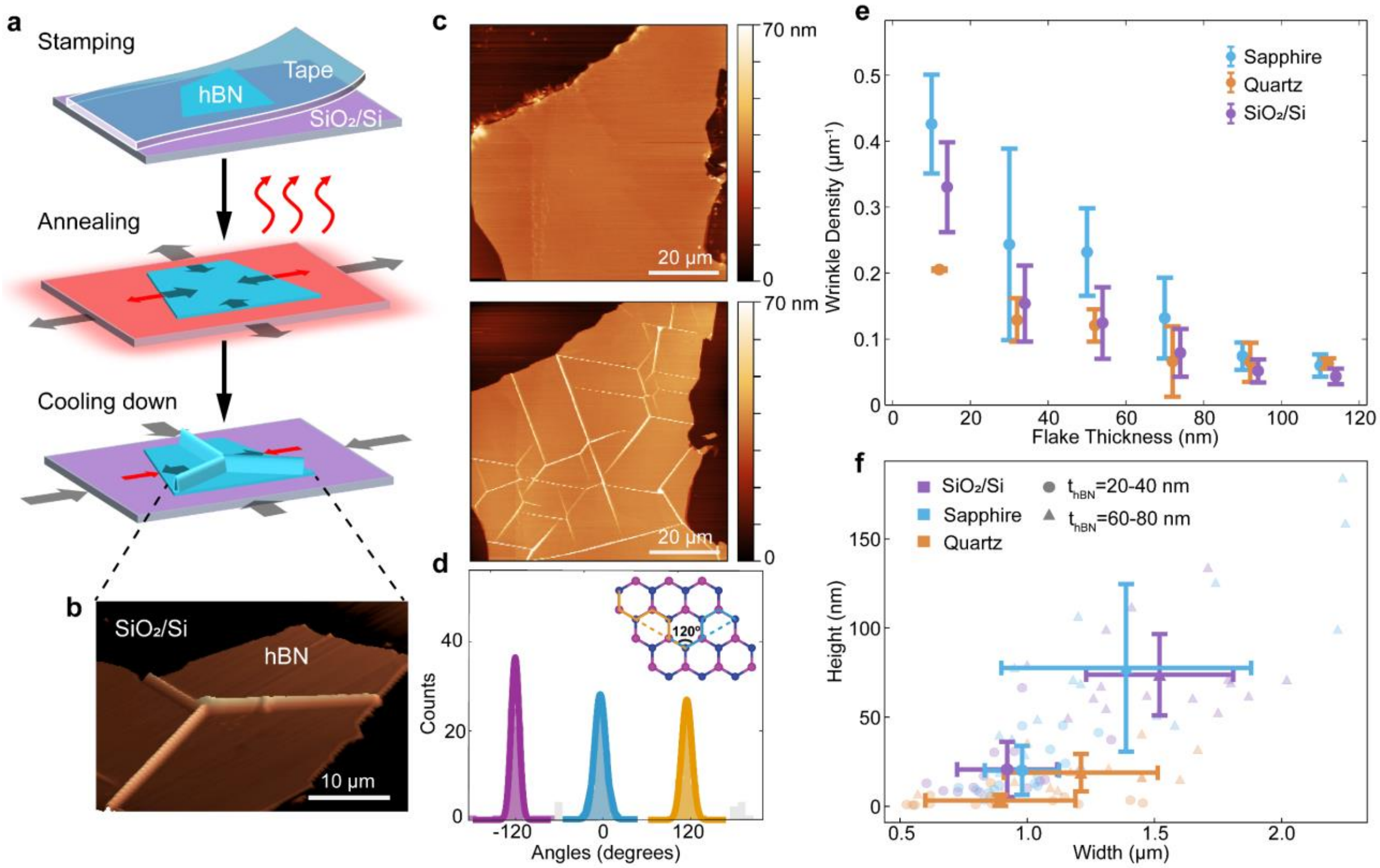

**Figure 1**. **Wrinkle formation and morphology design space in multilayer hBN.** (a) Schematic illustrating the wrinkle formation mechanism in hBN. (b) 3D AFM topography image showing a three-way wrinkle junction in an hBN flake. (c) AFM image of the hBN flake on $SiO_2$/Si before (top) and after (bottom) the formation of wrinkles by annealing. (d) Angular distribution of wrinkles, with an inset showing the 120° separation with respect to the hBN lattice. (e) Wrinkle density, quantified as total traced wrinkle length per unit area, versus hBN thickness for $SiO_2$/Si, sapphire, and quartz. (f) Wrinkle geometry statistics (height and width) versus thickness and substrate, showing nanometer scale vertical gaps with micron-scale lateral apertures and a strong dependence on substrate coupling and thickness.

Wrinkle cross sections are often approximately triangular with sharp curvature near the crest and trough (Figure 1b). Across the explored parameter space, wrinkle widths are in the micron range, while heights span from the sub-10 nm scale to hundreds of nanometers (Figure 1f). Here, wrinkle width is defined as the base-to-base distance of the AFM cross sectional profile. The most pronounced wrinkles are observed for thicker flakes on sapphire and $SiO_2$/Si, where larger mismatch and more efficient strain transfer can produce heights above 150 nm and widths up to about 2 μm. In contrast, thin hBN flakes on quartz reproducibly yield low amplitude nanowrinkles with heights below 50 nm. In the data for quartz substartes, approximately half of the measured wrinkles (~53%) fall in the near flat regime around the < 2 nm height scale, which is particularly attractive for subsequent liquid inflitration experiments because it provides nanometer scale vertical gaps while preserving wide optical access through the transparent substrate.

Although nominal thermal expansion mismatch provides a useful first order predictor, Figure 1 also suggests that additional factors modulate the effective strain transfer. For example, quartz and thermally grown $SiO_2$ can have comparable nominal expansion coefficients depending on the temperature range considered, yet quartz produces markedly smaller wrinkles and a weaker thickness dependence. This behavior is consistent with differences in substrate stiffness, interfacial pinning, and local adhesion heterogeneity that can change how compressive strain is transmitted into the flake and how buckling instabilities nucleate [21]. In this context, our data are better interpreted as a statistically tunable design space rather than a regime of deterministic control.

Finally, the observed scaling trends of wrinkle width and height with thickness and effective strain are consistent with classical thin film wrinkling mechanics. In classical wrinkling frameworks, the characteristic lateral length scale increases with film thickness and depends on the effective stiffness contrast, while the wrinkle amplitude increases with compressive strain beyond a critical threshold [22, 23]. Such relations have been broadly adopted to rationalize wrinkling in 2D materials, where thickness, substrate coupling, and interfacial interactions jointly define the emergent morphology distributions [20, 24]. The absence of obvious delamination and the coherent wrinkling of thick flakes are consistent with strong interlayer coupling in multilayer hBN [20]. Together, Figure 1 establishes that annealing induced wrinkling offers a lithography-free route to generate long range, surface integrated, channel-like pathways with nanometer scale vertical gaps and micron-scale lateral apertures, and that these features can be tuned statistically via substrate choice and flake thickness.

**Strain distribution along wrinkle-defined pathways and its optical signatures**

High temperature annealing produces an extended network of wrinkle ridges in multilayer hexagonal boron nitride (hBN), forming long, channel like pathways across the flake surface. Beyond providing geometric confinement through a sub-100 nm vertical gap, these ridges concentrate curvature and redistribute in plane stress within the multilayer sheet. As a result, a single wrinkle is not mechanically uniform; instead, the local strain varies across the wrinkle cross section, from the crest through the adjacent flanks to the trough regions. In the context of this work, resolving this geometry linked strain landscape is important for two reasons. First, it provides an intrinsic and label free handle to locate and register wrinkle defined pathways prior to liquid access and tracer experiments. Second, it explains why channels with different morphologies behave differently, so we focus on statistical trends rather than trying to deterministically control the exact channel geometry.

To resolve cross sectional strain partitioning by confocal Raman mapping (488 nm excitation, NA = 0.9; step size 0.1 μm), we analyze relatively wide, high amplitude wrinkle ridges whose dimensions are better matched to the sub-micron probe volume. This choice reduces diffraction limited spatial averaging that would otherwise wash out crest to trough contrast in ultrashallow nanowrinkles; for our optical configuration, the diffraction limited lateral resolution is on the order of $\delta \approx \frac{0.61\lambda}{NA} \approx 0.33\ \mu m (\lambda = 488\ \mathrm{nm}, \mathrm{NA} = 0.9)$ [25, 26]. Importantly, these high amplitude ridge show a qualitatively similar, ridge-like (often approximately triangular) cross section to the lower-amplitude wrinkles in Figure 1. We therefore use them as a representative geometry to extract the curvature-linked strain signature of wrinkle-defined channels.

Figure 2 combines AFM topography and Raman mapping to link wrinkle geometry to strain signatures. The AFM image (Figure 2a) shows representative ridge pathways and branching regions. Height profiles extracted across and along a ridge (Figure 2b) confirm the characteristic ridge like cross section and the continuity of the pathway over the measured length. The along ridge profile exhibits small point to point fluctuations that mainly arise from finite scan sampling and measurement noise, while the ridge height remains continuous along the measured segment, supporting the interpretation of wrinkles as extended pathways rather than isolated buckles.

Raman mapping of the hBN $E_{2g}$ peak of hBN, located at 1365.2 $cm^{-1}$ in the unstrained state, provides a direct readout of the spatial strain distribution associated with this geometry. In the Raman peak position map (Figure 2c), the $E_{2g}$ wavenumber varies systematically across the wrinkle. Relative to nearby flat regions, higher wavenumbers are observed near the ridge crest, whereas lower wavenumbers appear toward the ridge troughs. The corresponding line profile (Figure 2d) confirms this cross-sectional partitioning. Such a pattern is consistent with curvature driven redistribution of stress in a bent multilayer sheet, in which the local mechanical state differs between the crest region and the surrounding areas, a common outcome in strain engineering of 2D materials [24]. Because the Raman probe averages over a finite optical spot size, the extracted peak positions represent local averages within that optical volume, which becomes relevant for narrow wrinkles where crest and trough contributions can partially overlap.

Branching regions can further increase the spatial variability of the strain field because multiple ridge segments meet under geometric and boundary constraints. Prior simulations of wrinkle networks in supported hexagonal lattices predict alternating tensile and compressive regions distributed along and between intersecting ridges, with enhanced gradients near the intersection core [28]. While the present work does not rely on assigning a unique strain state to each branch, the Raman contrast supports the broader picture that wrinkle networks host spatially varying local environments along connected pathways rather than a single uniform strain state.

Although we primarily focus here on geometry-dependent strain, the same strain landscape can manifest as wavelength dependent optical contrast along wrinkles. Consistent with this expectation, photoluminescence (PL) maps acquired on representative wrinkles show that emission intensity in different spectral windows can highlight different parts of the wrinkle geometry, indicating that local environments associated with the crest and troughs can favor different radiative channels (in Figure S2). In wide bandgap hBN, optically active defect states can be perturbed by local strain and strain gradients, which can contribute to spatially heterogeneous optical response along a single wrinkle [27]. Related work in atomically thin semiconductors has also shown that spatially patterned local environments can impose nonuniform strain that modifies the band structure, thereby support large-scale arrays of optically active emitters [28]. Detailed discussion of defect and emitter physics is therefore provided in the Supporting Information, while here we emphasize that these optical signatures provide a practical way to optically address and register wrinkle segments prior to liquid access and biomolecular tracer experiments.

Overall, Figure 2 establishes that wrinkle defined pathways exhibit a reproducible, geometry linked strain signature that is directly accessible through Raman mapping and that correlates with wavelength dependent optical contrast. This combination of mechanical geometry and optical readout underpins the optically addressable character of the platform while also highlighting intrinsic heterogeneity, which motivates tuning of the morphology in subsequent sections.

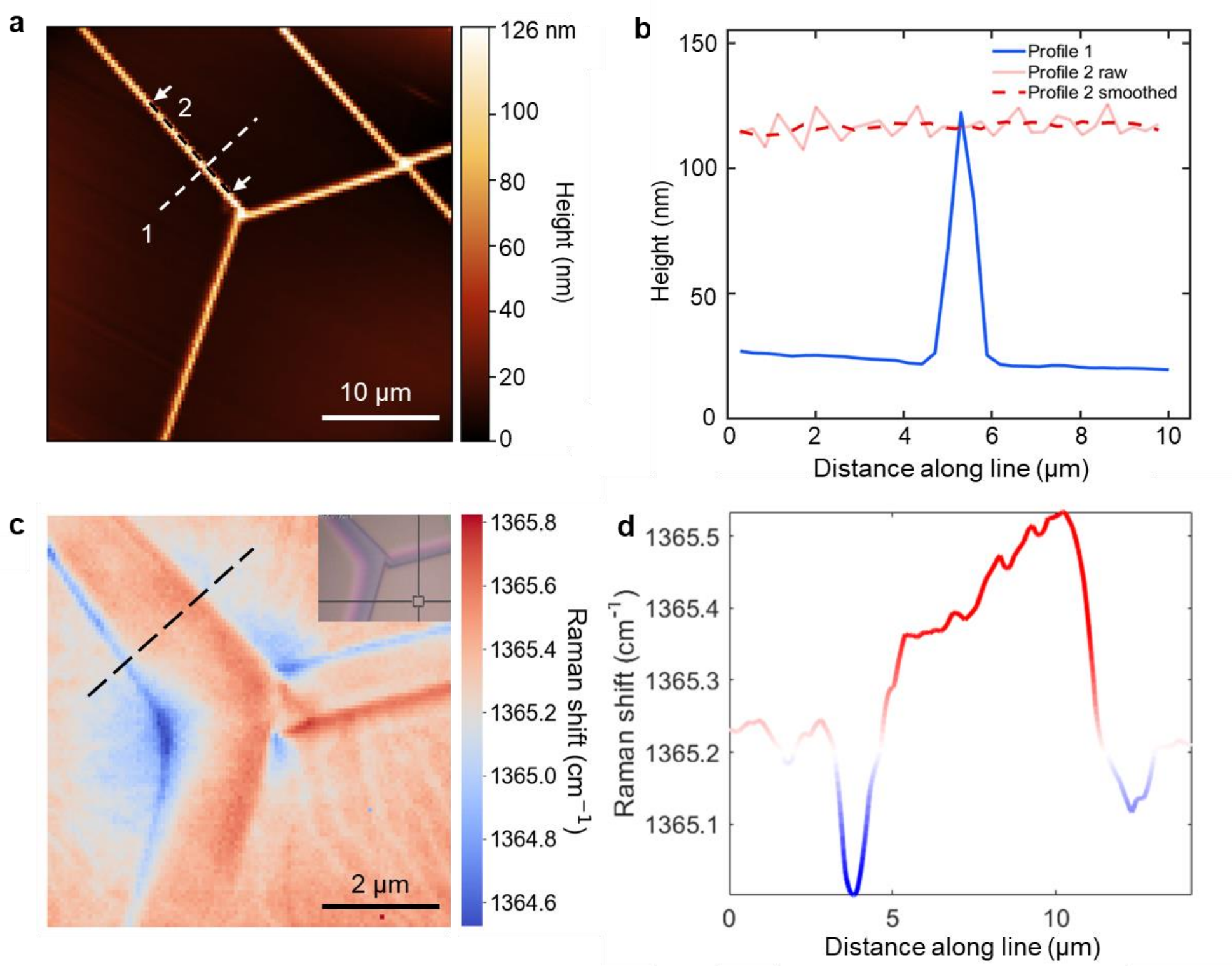

**Figure 2. AFM topography and Raman strain mapping of wrinkle-defined pathways in hBN. (a)** Representative AFM topography of annealing-induced wrinkles and branching junctions, illustrating continuous, surface-integrated pathways. **(b)** AFM height profiles extracted along the lines marked in (a): Profile 1 crosses a representative ridge to quantify ridge height and base-to-base width, and Profile 2 follows the ridge direction to illustrate height continuity along the pathway (white arrows mark the start and end of Profile 2). **(c)** Raman map of the hBN $E_{2g}$ peak position for a comparable wrinkle geometry, highlighting strain variations associated with the wrinkle morphology. (d) Raman $E_{2g}$ peak position extracted along the line marked in (c), showing the peak shifts across the wrinkle. AFM and Raman data were acquired on different flakes prepared under identical conditions with comparable thickness (∼100 nm) and consistent wrinkle morphology.

## Liquid filling of hBN nanochannels

Nanowrinkles not only provide tunable platforms for localized emission but also offer potential for nanofluidics research by confining liquids in nanochannels, opening up the route for combined biomolecule transport and optical based sensing in physiological conditions. To obtain a direct, real-space indication that liquids can access wrinkle-defined cavities, we first performed time-resolved optical imaging upon liquid contact using isopropanol (IPA) (Figure 3a). IPA was chosen here to facilitate capillary access and improve spreading at the flake surface, enabling the evolution of the optical contrast along the wrinkle pathways to be captured without specialized high-speed imaging. In the optical frames, the IPA droplet boundary on the terrace is outlined (yellow dashed line), while the wrinkle pathways (1 and 2) show a time-dependent contrast change that propagates along the channel direction (white arrows).

To quantify the observable propagation, we defined a reference point x=0 at the first resolvable contact location and tracked two quantities as a function of time (Figure 3b): (i) the apparent "liquid interface inside wrinkle", defined as the leading position of the contrast front along the selected wrinkle pathway, and (ii) the "droplet interface", defined as the position of the droplet boundary on the terrace relative to the same reference. The contrast front inside the wrinkle advances rapidly and reaches an approximately constant value (denoted as ∼saturated) within ∼0.6 s, whereas the droplet boundary continues to relax/spread more gradually. Importantly, the earliest stage of filling (approximately the first 0–0.2 s) is too fast to be resolved with our frame rate; therefore, the initial jump in the measured front position reflects a capillary-dominated liquid infiltration event that occurs between frames rather than a continuously sampled trajectory. This behaviour is consistent with prior reports that capillary infiltration in sub-micron conduits can proceed on sub-second timescales and may require high-speed imaging to fully resolve.[29]

Because the optical contrast method relies on a resolvable contrast front, we preferentially used relatively large wrinkles for visualization (typical width ~1 μm, height up to ~200 nm in Figure 3c), whereas smaller (~70 nm-scale) wrinkles are expected to fill even more rapidly and therefore appear "instantaneously filled" under standard video acquisition. Notably, besides contrast changes, some pathways also exhibit a subtle morphology-associated contrast evolution (pathway 2), suggesting that local meniscus formation and/or wrinkle deformation can accompany infiltration in larger wrinkles.

As an additional qualitative indicator of transport inside the confined pathway, we observed the motion of a sparse bright particle within a selected wrinkle after IPA contact under 532 nm laser excitation (Figure 3d). Here the particle is not an intentionally introduced tracer but an incidental particulate impurity/debris that is transported by the moving liquid within the channel. The sequential frames show a net displacement along the wrinkle axis (arrows), supporting the interpretation that the wrinkle channel can host mobile contents under droplet-driven capillary conditions. Aside from the molecular signal and the immobile reference, other spots arise from incidental impurities; they can move with in-channel flow and may quench over time. We thus demonstrate that the platform lends itself to direct optical readout of liquid access and in-channel motion.

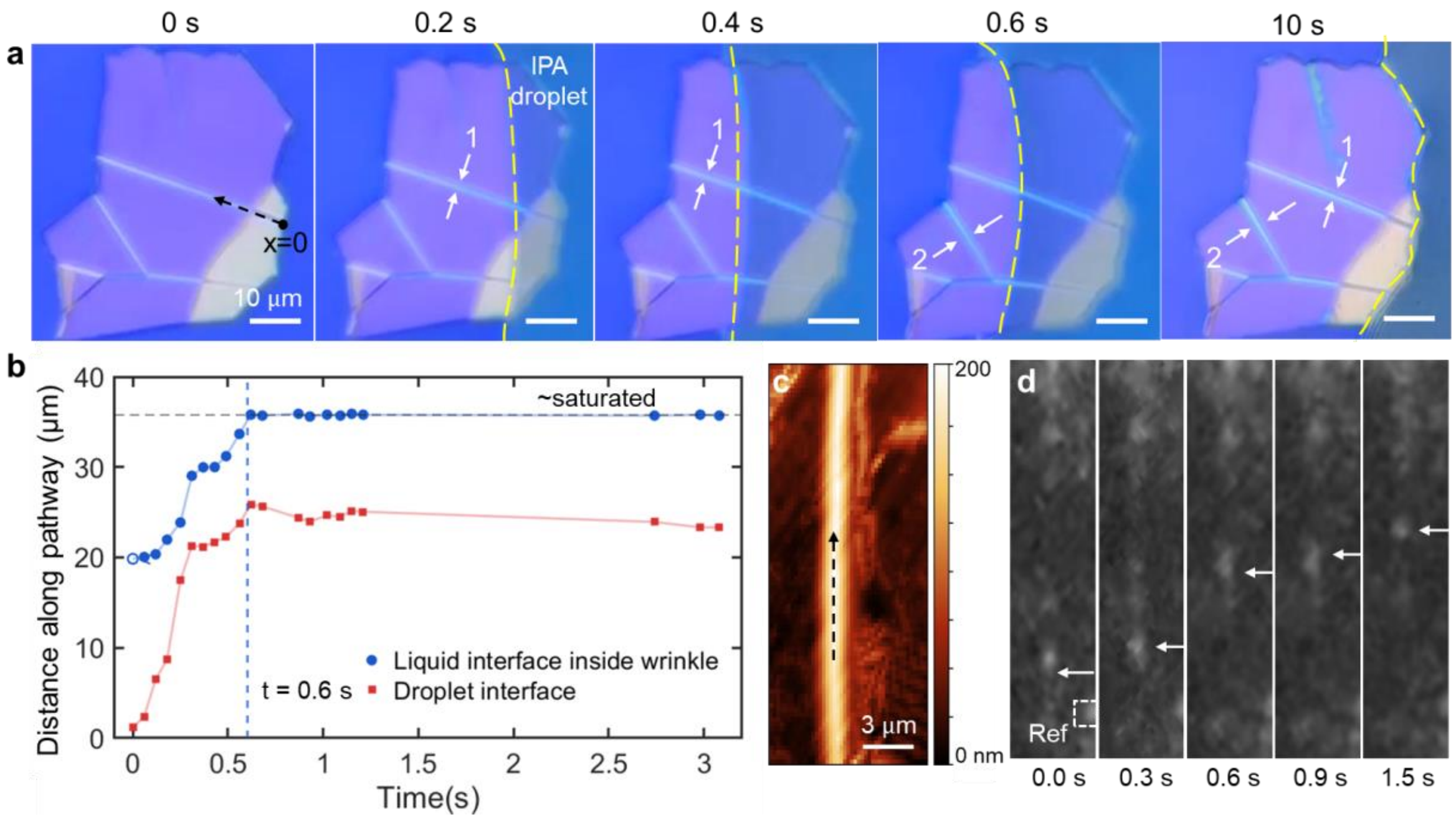

**Figure 3. Direct visualization of liquid infiltration and particle transport in wrinkle-defined hBN nanochannels. (a)** Time-sequence optical micrographs upon IPA droplet-contact, showing a contrast front propagating along a wrinkle (white arrows). The dashed yellow line indicates the droplet boundary. Two representative wrinkles are highlighted: wrinkle 1 shows a pronounced pathway-confined contrast change, whereas wrinkle 2 exhibits both contrast and apparent morphological evolution during the wetting process. The coordinate origin (x = 0) is defined at the droplet–wrinkle contact point used for distance quantification. Scale bars: 10 μm. **(b)** Positions of the apparent liquid interface inside the wrinkle (blue circles; defined as the leading edge of the contrast front along the pathway) and the droplet interface on the terrace (red squares) as a function of time after contact. The liquid front inside the wrinkle rapidly reaches a plateau (~saturated), while the droplet boundary continues to relax/advance on a longer timescale. The vertical dashed line marks the time at which the wrinkle signal becomes saturated within the temporal resolution of the optical recording. **(c)** AFM topography of a representative wrinkle segment used for particle-motion analysis. The dashed arrow indicates the pathway coordinate. Scale bar: 3 μm. **(d)** Time-resolved optical imaging showing the displacement of a particle transported within the wrinkle pathway (arrows). A stationary reference region (“Ref”) is used for image registration to exclude global drift.

To investigate whether wrinkles can effectively trap fluids beyond this direct optical evidence, we performed scanning KPFM[30] and Raman spectroscopy[31] before and after liquid infiltration. As a first step, we probed the dielectric response of water confined inside the wrinkles[32] using amplitude-modulation (AM) scanning dielectric microscopy[33] based on electrostatic force detection with an AFM. In this configuration, a single low-frequency AC voltage $V_{AC}\cos(\omega t)$ was applied between the conductive tip and the 50 nm ITO-coated quartz substrate, and the resulting electrostatic force $F_{el}(t)$ was analyzed at the modulation and second-harmonic frequencies:[30],[33]

$$F_{el}(t) = \frac{1}{2}\frac{dC}{dz}(V_{DC} - V_{CPD})^2 + \frac{dC}{dz}(V_{DC} - V_{CPD})V_{AC}\cos(\omega t) + \frac{1}{4}\frac{dC}{dz}V_{AC}^2\cos(2\omega t)$$

where C is the local tip–sample capacitance, z is the tip-sample separation, $V_{DC}$ is the applied DC bias, $V_{CPD}$ is the contact potential difference, and $V_{AC}$ is the applied AC bias. This single-pass AM mode measures the first derivative of the capacitance $\frac{dC}{dz}$ by focusing on the second harmonic component $\frac{1}{4}\frac{dC}{dz}V_{AC}^2\cos(2\omega t)$, which is independent of the potential term and directly proportional to $\frac{dC}{dz}$. This isolates the contribution from changes in local capacitance caused by the wrinkle interior, allowing us to detect the presence of confined water or other fluids inside the wrinkle nanochannels (Figure 4a).

We selected ~7 nm thick hBN flakes containing wrinkles with heights ranging from 20 nm to 60 nm. Milli-Q water was drop-cast onto the flake for 1 min, followed by thorough drying with a nitrogen gun to remove residual surface water, thereby ensuring that the AFM tip remained isolated from liquid contamination. Figures 4b to 4d show AFM topography and corresponding line profiles of three wrinkles before and after water infiltration. After water filling, the wrinkle heights increased slightly by 2–5 nm, indicating that water was encapsulated, slightly expanding the wrinkles vertically and reducing their sagging.

Notably, infiltration did not occur uniformly across the entire network. The wrinkles formed by high temperature annealing often intersect along the armchair crystallographic direction, creating characteristic three-way junctions that interconnect individual wrinkles (Figure 1a). Such junctions serve as natural nodal points that connect neighboring channels, enabling water to enter wrinkles near the flake edge and subsequently propagate into adjacent wrinkles through the junction. This interconnected geometry provides a percolating nanochannel network capable of guiding liquids and potentially different biomolecules from separate entry paths to a common junction region.

Capillary effects play a decisive role in determining which wrinkles are filled. Wrinkles connected to the external environment at both ends form continuous channels and are readily filled by capillary forces within short liquid exposure times. In contrast, wrinkles with only one open end tend to retain trapped air, creating pressure barriers that inhibit infiltration. Very high or partially collapsed wrinkles may further create internal blockages, isolating air pockets and preventing liquid entry. Such capillary-driven selectivity of filling has been widely observed in confined micro- and nanochannels.[34],[35]

KPFM images before and after filling are shown in Figures 4e-g. After Milli-Q water infiltration, the dielectric contrast became more negative, consistent with the increase in the relative dielectric constant from ~0 (air) to ~80 (Milli-Q water).[36] This increase in permittivity enhances the local capacitance gradient ($|\frac{dC}{dz}|$), thereby producing a stronger second-harmonic electrostatic signal in our measurements. Quantitatively, the capacitance-gradient amplitude increased by about 2.4 times ($|\frac{dC}{dz}|_{after}/|\frac{dC}{dz}|_{before} \approx 2.4$), indicating that water infiltration significantly amplified the dielectric contrast between the wrinkle interiors and surrounding regions. The overall baseline also increased, reflecting the higher effective permittivity of the tip–sample junction and the contribution of a thin adsorbed water layer that enhances background capacitance. The observed contrast change, particularly at the three-junction sites, clearly demonstrates that water was successfully transported between interconnected wrinkles and retained within the confined regions.

Interestingly, while some wrinkle shapes changed slightly after Milli-Q water filling, the three-way wrinkle junctions remained unchanged, in agreement with simulations showing that these junctions represent mechanically stable configurations under fluid confinement.[37] These stable nodal junctions therefore represent robust structural units for continuous nanofluidic networks, potentially enabling controllable biomolecule mixing or reaction sites at the intersection points.

As shown in Figure S3, replacing the confined Milli-Q water with a 1× TAE buffer containing 10 mM $MgCl_2$ (commonly used for DNA imaging experiments) resulted in the $\frac{dC}{dz}$ contrast becoming less negative. This is consistent with the reduced dielectric constant of the buffer compared to Milli-Q water, owing to ion–dipole interactions and reduced water mobility in electrolyte solutions.[38] The ability to distinguish dielectric changes between Milli-Q water and electrolyte solutions in individual wrinkles demonstrates the sensitivity of our approach and lays the groundwork for optical–electrical hybrid sensing of liquid mixtures and even potentially biomolecules within wrinkle-based hBN nanochannels.

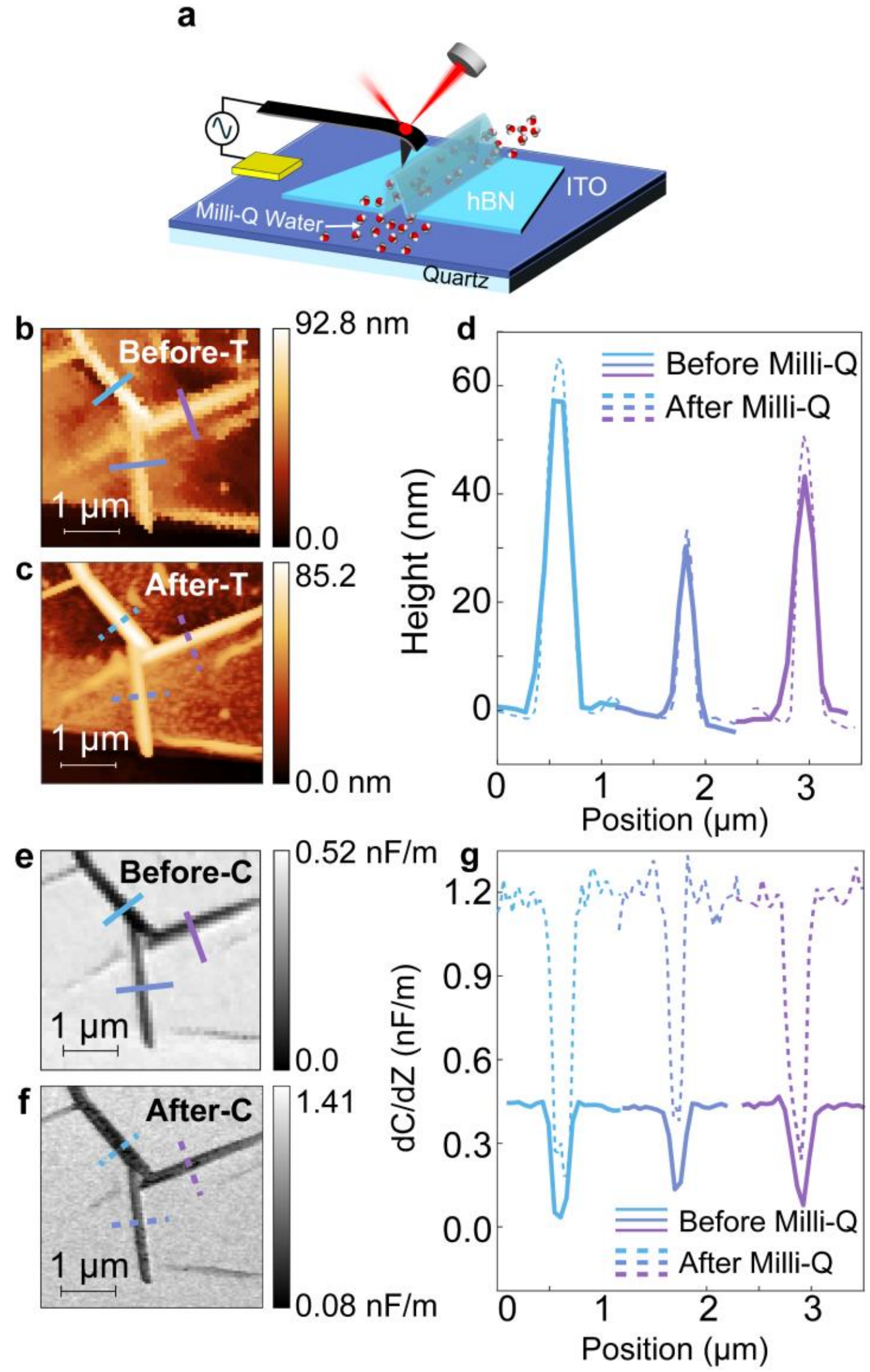

**Figure 4. Dielectric imaging of confined water inside the wrinkles. (a)** Schematic of the KPFM setup and sample. Topographic AFM images of the wrinkles **(b)** before and **(c)** after filling with Milli-Q water. **(d)** Corresponding topographic profiles of three wrinkles measured before (solid lines) and after Milli-Q water filling (dashed lines). Capacitance gradient (dC/dz) images of the wrinkles **(e)** before and **(f)** after Milli-Q water filling. **(g)** Corresponding capacitance gradient profile of three wrinkles obtained before (solid lines) and after Milli-Q water filling (dashed lines)

To complement the variation in dielectric properties with a spectroscopic signature of confined water, we further performed Raman mapping to track water distribution in the wrinkle network (Figure 5), We carried out Raman mapping on nanowrinkles at two stages of the water infiltration experiment: firstly, at 2 minutes of water immersion followed by drying, and secondly, after 1 hour of immersion followed by drying. The corresponding Raman maps are presented in Figures 5a-c. A clear increase in the water-related Raman signal (3360 $cm^{-1}$) is observed with longer immersion time (from 2 min to 60 min), indicating a higher water content trapped inside the wrinkles as a function of time. This is consistent with the principle that Raman peak intensity is proportional to the amount of material present within the probed volume, in our case, the amount of residual water confined in the wrinkles.

Detailed inspection of the Raman maps reveals that the strongest water signal often appears at the junction region where three wrinkles converge (Figure 5b). In some cases, the central meeting point is not completely open but remains mechanically connected to the three surrounding wrinkles, allowing liquid to pass through even when the core appears partially closed. This observation further supports the notion of a continuous nanofluidic network formed by interconnected wrinkles, in agreement with the KPFM results.

We further verified that the trapped substance is water by comparing the Raman spectrum of Milli-Q water with that collected from the wrinkle regions (Figure 5d). The characteristic broad OH-stretch band of water (3100–3600 $cm^{-1}$) appears exclusively within the wrinkles and not in nearby flat areas. These results demonstrate that the wrinkle junctions act as stable nodal points for liquid transport and accumulation, while the surrounding wrinkles serve as capillary channels feeding these junctions. This behavior is consistent with previous studies showing that confined nanoscale geometries such as cracks or wrinkles can function as capillary conduits that promote fluid infiltration and retention.[39]

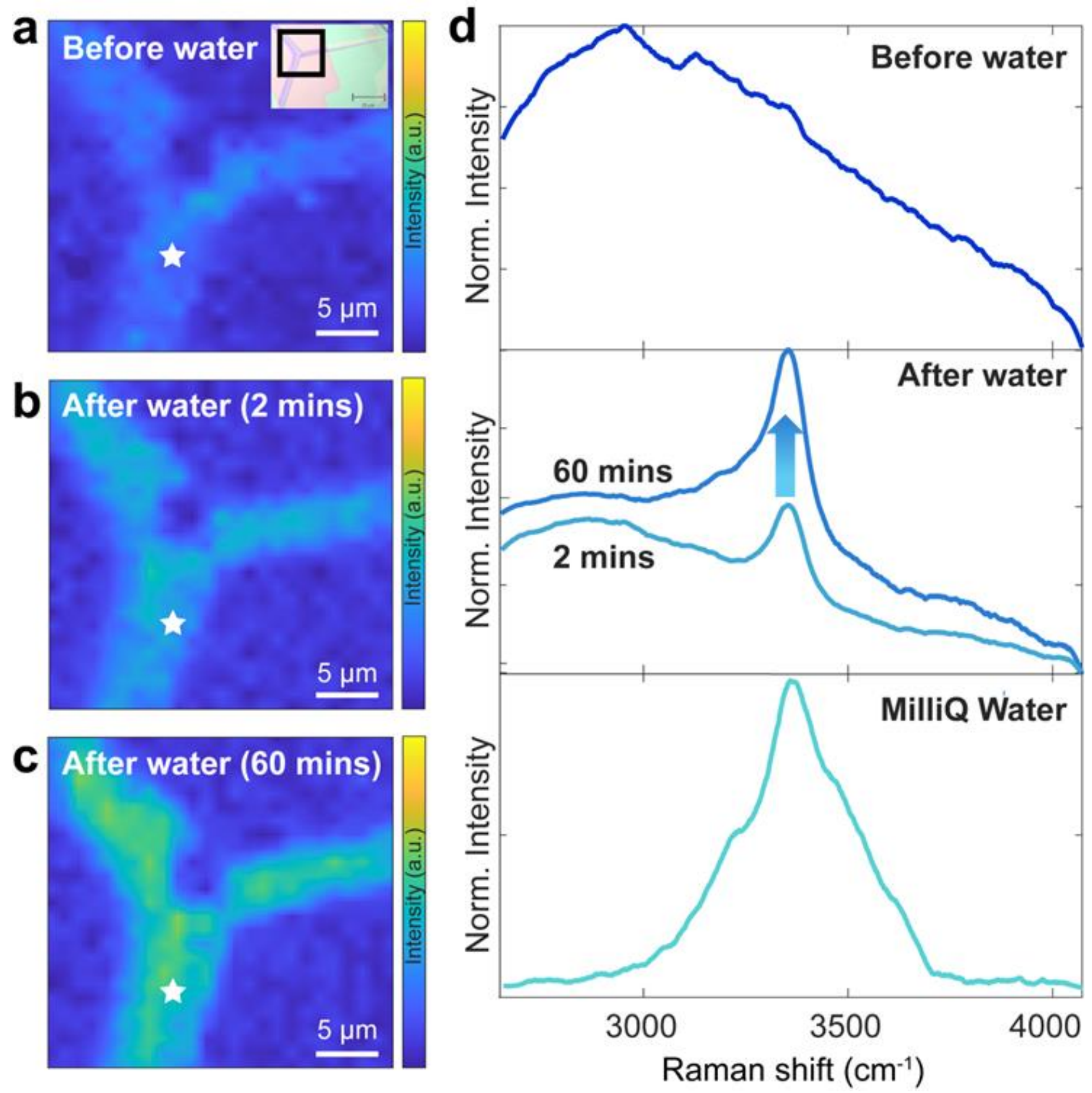

**Figure 5. Raman measurement of confined water inside the wrinkles.** (a–c) Raman intensity maps of the wrinkle region before Milli-Q water exposure (a), after 2 min (b), and after 60 min (c) of Milli-Q water induction, based on the characteristic Raman peak of Milli-Q water. (d) Corresponding Raman spectra collected at the wrinkle location indicated by the star symbol in (a–c), shown before and after water exposure for 2 min and 60 min, together with the reference spectrum of pure Milli-Q water.

### Fluorescence-based detection of confined DNA in wrinkle-induced hBN/graphene nanochannels

After confirming liquid and particles confinement inside wrinkle channels, we subsequently introduced ATTO647N-labeled single-stranded DNA (ssDNA) as a fluorescent molecular probe to demonstrate that wrinkles can act as nanochannels that enable confinement and imaging of analytes in solution. Specifically, we evaluate (i) DNA loading from an aqueous droplet, (ii) localization and retention of a detectable fluorescent signal along the channel geometry, and (iii) selective optical readout under physiologically relevant buffer conditions.

Following the droplet-contact loading procedure used in the IPA experiments (Figure 3), we introduced an ATTO647N–ssDNA solution and allowed it to spread across the flake such that the droplet front contacted the wrinkle network. Fluorescence imaging showed enhanced signals at wrinkle locations, suggesting preferential accumulation along the wrinkle-defined pathways (Figure S4). However, AFM characterization after drying (Figure S4f) indicates that the observed fluorescence cannot be unambiguously assigned to ssDNA inside the nanochannel, because ssDNA can also adsorb on the wrinkle crest and the surrounding hBN surface. Therefore, this observation is treated as qualitative evidence of wrinkle-associated localization rather than definitive proof of in-channel confinement. Such adsorption is likely promoted by native defects that act as trapping sites, slowing or immobilizing DNA molecules.[40]

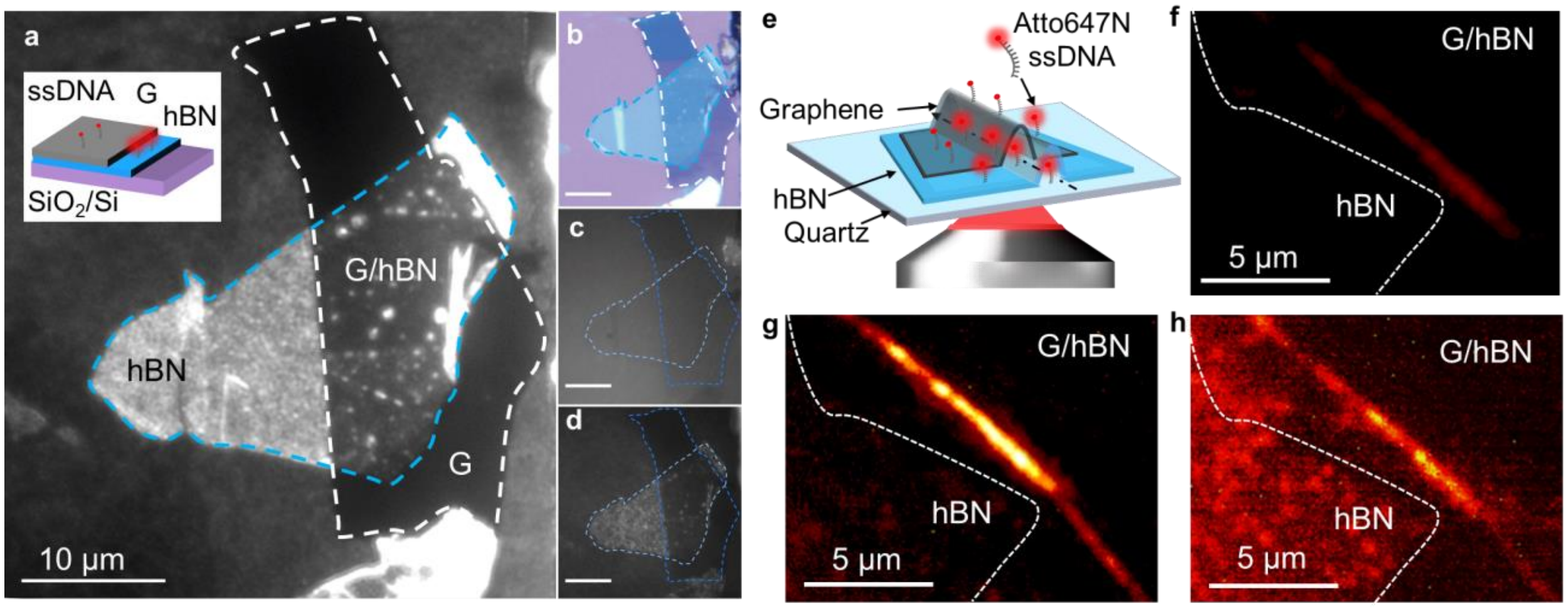

**Figure 6. hBN/graphene nanochannels for selective fluorescence detection of confined biomolecules.** (a) Epifluorescence image of 100 pmol ATTO647N ssDNA on a graphene/hBN stack 1 min after flushing in the biomolecules. Inset: schematic of the vertical van der Waals structure. (b) Wide-field optical image of the stack before exposure to Milli-Q water and biomolecules. (c,d) Epifluorescence images in Milli-Q water before and immediately after the addition of ATTO647N ssDNA respectively. (e) Schematic of the hBN/graphene nanochannel for ssDNA confinement. (f) Epifluorescence image in air of the hBN/graphene stack. (g,h)

Epifluorescence image 5 min after the addition of 10 pmol ssDNA addition and after a 10h incubation. The excitation wavelength is 640 nm.

To suppress background fluorescence from surface-adsorbed ssDNA, a thin graphene layer was transferred onto the wrinkled hBN film, forming a vertical hBN/graphene heterostructure designed to mask surface background while preserving emission from molecules residing deeper in the wrinkle cavity (Figure 6). Graphene quenches nearby fluorophores through non-radiative energy transfer,[17] an effect also observed in DNA–graphene hybrid systems[41] and patterned hBN/graphene heterostructures used as fluorescence masks.[42] In this configuration, the quenching efficiency decays rapidly with distance: when the hBN thickness exceeds the characteristic distance for 50% excitation energy transfer (~11 nm)[17], the hBN dielectric layer sufficiently separates the fluorophores from graphene, allowing their emission to recover. Consequently, only fluorophores located within regions where the hBN layer is thicker than this threshold, for example closer to the base of the wrinkle channels, remain optically detectable.

To verify the quenching performance of graphene, we first examined a flat hBN/graphene heterostructure in which a thin graphene layer ($t_{Graphene}$ ~ 4.5 nm) was transferred on top of an hBN flake ($t_{hBN}$ ~ 7.5 nm) (Figures 6a–d). Fluorescence imaging (Figure 6a) reveals that the hBN region exhibits uniform bright emission due to ssDNA adsorption, whereas the region covered by graphene shows complete quenching. The corresponding optical image (Figure 6b) highlights the hBN flake (blue) and the overlying graphene layer (purple). Before ssDNA deposition (Figure 6c), no fluorescence background is visible, while during ssDNA addition (Figure 6d), transient intensity fluctuations appear as molecules interact with the surface. After rinsing and drying, the background fluorescence from surface-adsorbed ssDNA disappears, confirming that the top graphene efficiently quenches nearby fluorophores through non-radiative energy transfer, thereby validating this configuration as a quenching control. Occasional residual bright spots in the overlap region likely arise from local interfacial inhomogeneities (e.g., trapped bubbles or slight delamination), which locally reduce quenching efficiency.

Building on these stacked-layer experiments, we then annealed an hBN/graphene stack ($t_{Graphene}$ ~ 10 nm; $t_{hBN}$ ~ 68 nm) to induce wrinkle formation and introduced ssDNA into the resulting nanochannels (Figures 6e–h). Before ssDNA deposition (Figure 6f), weak emission could occasionally be observed at the strained wrinkle sites, corresponding to pre-existing, strain-activated hBN emitters. Upon droplet-contact and subsequent spreading of the ssDNA solution, the wrinkle channels become loaded predominantly by capillary forces. Consistent with the capillary-dominated loading observed in Figure 3, the initial filling/localization can occur on sub-second timescales in nanochannels and may not be resolved at standard camera frame rates. Even at the minimum syringe-pump rate (3.076 μL $h^{-1}$, NE-1000), the flow could not be stabilized, confirming that spontaneous capillary forces dominate. After ssDNA immersion and drying (Figure 6g), clear fluorescence appeared selectively inside the wrinkles, while nearby ssDNA on graphene was efficiently quenched.

The confined ssDNA solution remains stable for approximately 10 hours under ambient conditions due to strong capillary pressure in the nanoscale confinement that suppresses evaporation. Additionally, the weak van der Waals interactions at the liquid–hBN interface help to stabilize the thin film, and limit air exchange through the narrow channel ends. The ssDNA stays visibly confined within the channels, and as evaporation gradually proceeds, partial drying near the wrinkle openings likely drives the remaining ssDNA to redistribute toward more confined regions of the channels as shown in Figure 6h.

In addition to these successful loading cases, we note a common failure mode in which liquid/DNA does not propagate into the wrinkle interior due to trapped air (interfacial bubbles) or locally blocked / dead-end geometries. In Supplementary Figure S5a–c, a bubble-containing or blocked wrinkle do not develop a continuous fluorescence signal along the channel, whereas an open wrinkle exhibits the expected continuous localization (Figure S5d–f). The observations that DNA readout requires an free capillary pathway and that trapped air can act as a kinetic barrier are consistent with known backpressure and bubble effects in confined capillary filling[43]. Practically, degassing and/or alcohol (IPA/ethanol) pre-wetting are standard routes to reduce bubble nucleation/persistence and improve filling probability in micro/nanofluidic handling, offering a straightforward strategy to enhance loading reproducibility for wrinkle networks[44].

Control experiments using only buffer solution showed no detectable fluorescence, as evidenced in Figure S6, confirming that the buffer itself does not introduce background emission. Importantly, the native hBN emitters remained unchanged under buffer treatment, demonstrating that the wrinkle-induced channels are optically stable and that buffer immersion does not significantly activate or modify intrinsic hBN emission. In all experiments, ssDNA deposition was performed by immersion followed by drying to minimize background fluorescence from free molecules in solution. Overall, these results position the wrinkle-defined hBN/graphene heterostructure as a functional nanochannel platform that enables high contrast optical readout of biomolecules localized within the wrinkle geometry, with prospects in revealing biomolecule-biomolecule interactions and transport dynamics in confinement.

## Conclusion

Through a systematic study of thermally induced wrinkling in multilayer hBN on substrates with different thermal expansion mismatch, we identify how substrate choice and flake thickness jointly regulate wrinkle density, dominant morphology, and network connectivity. Rather than targeting fully deterministic channel dimensions, we frame wrinkle formation as a design space in which distinct regimes (isolated ridges, junctions, and percolating networks) can be accessed by tuning sample and fabrication parameters. Combined AFM and Raman mapping link wrinkle geometry to strain signatures, providing an optical handle for locating and comparing different morphologies.

Additionally, we show that wrinkle networks are accessible to aqueous liquids and can retain confined liquid over extended times (~hours). Time-resolved optical imaging upon droplet contact indicates rapid access and propagation of liquid along connected channels under capillary-dominated loading. Because the initial infiltration occurs on sub-second timescales that are faster than the camera frame rate, the apparent liquid "front" evolution mainly reflects droplet spreading and the expanding contact with the wrinkle network. To verify liquid presence within the wrinkle-defined cavity, we combine Raman mapping of the water OH-stretch band with capacitance-gradient (dC/dz) dielectric contrast measured using KPFM-based scanning dielectric microscopy; both signals co-localize with the wrinkle network after exposure and drying and persist on experimentally relevant timescales (>10 h). The agreement between (i) time-dependent optical access upon droplet-contact, (ii) spectroscopic detection of the water band by Raman mapping, and (iii) dielectric contrast changes by KPFM proves inflitration and long-term retention of confined liquid. While loading is not uniformly successful across all wrinkles, we address this variability by selecting channels and junctions that are free of obvious blockage or dead-end terminations, emphasizing that connectivity and interface condition are critical.

To enable bio-relevant optical readout, we integrate a graphene overlayer as a distance-dependent background-suppression interface, allowing preferential wide-field fluorescence localization of labeled DNA within the channels as a proof-of-concept application. At the same time, we identify key limitations, including nonspecific adsorption on hBN, transfer-induced interfacial inhomogeneities, and morphology heterogeneity. Future work can address these limitations by (i) implementing surface passivation or chemical functionalization to reduce adsorption, (ii) improving transfer uniformity and employing degassing/priming protocols (e.g., alcohol pre-wetting) to mitigate trapped-air failure modes, and (iii) adding quantitative transport metrics (pressure-driven flow, ionic conductance, or calibrated diffusion measurements) to connect morphology to nanofluidic performance. Overall, wrinkled hBN provides a simple, scalable route to achieve optically addressable planar nanofluidic confinements with prospects in nanoscale characterization of liquid mixtures and (bio)molecular analysis.

## Methods

### Sample Preparation

Multilayer hBN flakes were first mechanically exfoliated from bulk hBN crystals produced by high temperature and high-pressure synthesis (NIMS Japan) and then transferred onto $SiO_2$/Si wafer chips cleaned by ultrasonic sonication in acetone and IPA for 3 min and oxygen plasma cleaning for 5 min. Atomic force microscopy (AFM, Cypher, Asylum Research) was used to characterize the surface morphology and wrinkle structures of hBN flakes. For the systematic wrinkle characterization shown in Figure 1, the hBN flakes on different substrates were annealed in high vacuum ($\sim 7.4 \times 10^{-5}$ mbar) at 1000 °C for 1 h. For subsequent measurements (Figures 2–5), a shorter annealing procedure at 850 °C for 15 min in forming gas (95% $N_2$/5% $H_2$) was used to obtain similar wrinkle features more rapidly. Both annealing environments are known to promote strain relaxation and wrinkle formation in layered materials through out-of-plane buckling under compressive stress generated during cooling.[45]

### Raman and PL measurements

Raman and PL spectra were acquired with a Renishaw Raman setup. The hBN samples were excited by a 514 nm argon laser (MODU-LASER) with 0.15 mW power. The spectra were collected by an objective lens (Olympus 50×) with numerical aperture (NA) of 0.6 and a 1800 l/mm grating. The PL mapping was performed with the same setup and excitation laser, but with a higher resolution objective (Olympus 100x) with NA 0.9. For the PL maps, a laser scanning step of 0.5 µm and 1 µm was used.

**Kelvin Probe Force Microscopy (KPFM) measurement**

All measurements were performed on an Oxford Instruments/Asylum Research CypherAFM under ambient conditions. We used Bruker FMV-PT conductive probes, which feature a platinum-iridium (Pt/Ir) coated tip on an antimony (Sb) doped silicon cantilever. The cantilever have a nominal spring constant of 2.8 N/m, a resonant frequency of ~75 kHz, and a tip radius of ~25 nm. Single-pass KPFM was employed in Figure 4 and dual-pass KPFM in Figure S3 to characterize the sample's surface potential. The use of different modes was primarily dictated by practical considerations related to scan stability and compatibility with the liquid environment. In principle, both methods yield comparable dC/dz contrast, as confirmed by previous studies.[30, 46, 47] In the dual-pass configuration, the cantilever is excited near its resonance frequency, resulting in a higher signal-to-noise ratio, although the overall contrast remains consistent with that obtained in single-pass measurements.

**Single-Pass AM-FM Mode**

For high-resolution, simultaneous mapping of topography and surface potential, we employed Asylum's single-pass amplitude-modulation (AM) KPFM mode. In this configuration, the cantilever was driven at its mechanical resonance frequency for topography imaging, while a low-frequency AC voltage (5 kHz, 3 Vpp) was applied between the conductive tip and the sample. This single-frequency excitation induces a time-varying electrostatic force that modulates the cantilever oscillation amplitude. The electrostatic response was analysed at both the fundamental and second-harmonic frequencies, where the latter provides sensitivity to variations in the local capacitance gradient (dC/dz). Using the measured spring constant and detector responsivity, the 2ω signal was converted to dC/dz values according to Equation 1. The sample was first characterized in air and then remeasured after immersion in Milli-Q water for 1 min.

**Dual-Pass (Lift) Mode KPFM**

We also employed a dual-pass Kelvin Probe Force Microscopy (KPFM) technique, commonly known as Lift Mode. For each scan, the surface topography was first recorded using AC mode. In the subsequent interleave pass, the tip was lifted to a constant height of 30 nm above the recorded profile. During this lift pass, the mechanical excitation is switched off, and the tip is excited by a 3 Vpp AC voltage between tip and sample, oscillating at the mechanical resonance frequency (63.5 kHz) of the cantilever. A KPFM feedback loop adjusted the DC bias on the tip to nullify the first harmonic amplitude at 63.5 Hz. This recorded DC bias directly corresponds to the surface potential map, revealing distinct potential domains across the sample with typical values around 130.69 mV. The second harmonic again corresponds to the dC/dz signal.

The experimental workflow involved two consecutive measurements on the same sample area to compare the effects of Milli-Q water and a buffer solution. First, the sample was immersed in MilliQ water for 3 min, after which a KPFM scan was performed under ambient conditions. Subsequently, the same area was immersed in the buffer solution for 3 minutes and scanned again to map the resulting changes in surface potential.

ASSOCIATED CONTENT

**Supporting Information**

The following files are available free of charge.

AUTHOR INFORMATION

**Corresponding Author**

* Email: s.caneva@tudelft.nl

**Author Contributions**

The manuscript was written through contributions of all authors. All authors have given approval to the final version of the manuscript.

ACKNOWLEDGMENT

X.Y. acknowledges funding from the Chinese Scholarship Council (Scholarship No. 202108270002). S.C. acknowledges funding from the European Union's Horizon 2020 research and innovation program (ERC StG, SIMPHONICS, Project No. 101041486). All authors acknowledge K. Watanabe and T. Taniguchi from the National Institute of Materials Science (NIMS) for the bulk hBN crystals, and gratefully acknowledge P.G. Steeneken for valuable discussions and insightful comments during the final stage of manuscript preparation.

**Notes**

The authors declare no competing financial interests.

REFERENCES

[1] Whitesides, G. M. The origins and the future of microfluidics. *nature* **2006**, *442*, 368-373.
[2] Bocquet, L.; Charlaix, E. Nanofluidics, from bulk to interfaces. *Chemical Society Reviews* **2010**, *39*, 1073-1095.
[3] Schoch, R. B.;Han, J.; Renaud, P. Transport phenomena in nanofluidics. *Reviews of modern physics* **2008**, *80*, 839-883.
[4] Mijatovic, D.;Eijkel, J. C.; Van den Berg, A. Technologies for nanofluidic systems: top-down vs. bottom-up—a review. *Lab on a Chip* **2005**, *5*, 492-500.
[5] Dean, C. R.;Young, A. F.;Meric, I.;Lee, C.;Wang, L.;Sorgenfrei, S.;Watanabe, K.;Taniguchi, T.;Kim, P.; Shepard, K. L. Boron nitride substrates for high-quality graphene electronics. *Nature nanotechnology* **2010**, *5*, 722-726.
[6] Radha, B.;Esfandiar, A.;Wang, F.;Rooney, A.;Gopinadhan, K.;Keerthi, A.;Mishchenko, A.;Janardanan, A.;Blake, P.; Fumagalli, L. Molecular transport through capillaries made with atomic-scale precision. *Nature* **2016**, *538*, 222-225.
[7] Keerthi, A.;Geim, A. K.;Janardanan, A.;Rooney, A. P.;Esfandiar, A.;Hu, S.;Dar, S. A.;Grigorieva, I. V.;Haigh, S. J.; Wang, F. Ballistic molecular transport through two-dimensional channels. *Nature* **2018**, *558*, 420-424.
[8] Vasu, K.;Prestat, E.;Abraham, J.;Dix, J.;Kashtiban, R. J.;Beheshtian, J.;Sloan, J.;Carbone, P.;Neek-Amal, M.; Haigh, S. Van der Waals pressure and its effect on trapped interlayer molecules. *Nature communications* **2016**, *7*, 12168.
[9] Keerthi, A.;Goutham, S.;You, Y.;Iamprasertkun, P.;Dryfe, R. A.;Geim, A. K.; Radha, B. Water friction in nanofluidic channels made from two-dimensional crystals. *Nature communications* **2021**, *12*, 3092.
[10] Zhang, Z.;Gudarzi, M. M.;Mao, J.;Wang, Z.;Bedran, Z.;Chen, C.;Nonahal, M.;Timokhin, I.;Mishchenko, A.; Yang, Q. Rapid fabrication of clean van der Waals nanochannels using Mask and Stack method. *arXiv preprint arXiv:2511.18480* **2025**.
[11] Genzer, J.; Groenewold, J. Soft matter with hard skin: From skin wrinkles to templating and material characterization. *Soft Matter* **2006**, *2*, 310-323.
[12] Chen, W.;Gui, X.;Yang, L.;Zhu, H.; Tang, Z. Wrinkling of two-dimensional materials: methods, properties and applications. *Nanoscale Horizons* **2019**, *4*, 291-320.
[13] Oliveira, C. K.;Gomes, E. F.;Prado, M. C.;Alencar, T. V.;Nascimento, R.;Malard, L. M.;Batista, R. J.;de Oliveira, A. B.;Chacham, H.; de Paula, A. M. Crystal-oriented wrinkles with origami-type junctions in few-layer hexagonal boron nitride. *Nano Research* **2015**, *8*, 1680-1688.
[14] Paszkowicz, W.;Pelka, J.;Knapp, M.;Szyszko, T.; Podsiadlo, S. Lattice parameters and anisotropic thermal expansion of hexagonal boron nitride in the 10–297.5 K temperature range. *Applied Physics A* **2002**, *75*, 431-435.
[15] Wang, R.;Souilamas, M.;Esfandiar, A.;Fabregas, R.;Benaglia, S.;Nevison-Andrews, H.;Yang, Q.;Normansell, J.;Ares, P.;Ferrari, G.;Principi, A.;Geim, A. K.; Fumagalli, L. In-plane dielectric constant and conductivity of confined water. *Nature* **2025**, *646*, 606-610.
[16] Gómez-Santos, G.; Stauber, T. Fluorescence quenching in graphene: A fundamental ruler and evidence for transverse plasmons. *Physical Review B—Condensed Matter and Materials Physics* **2011**, *84*, 165438.
[17] Yang, X.;Shin, D. H.;Yu, Z.;Watanabe, K.;Taniguchi, T.;Babenko, V.;Hofmann, S.; Caneva, S. Hexagonal Boron Nitride Spacers for Fluorescence Imaging of Biomolecules. *ChemNanoMat* **2024**, *10*, e202300592.

[18] Chen, L.;Elibol, K.;Cai, H.;Jiang, C.;Shi, W.;Chen, C.;Wang, H. S.;Wang, X.;Mu, X.; Li, C. Direct observation of layer-stacking and oriented wrinkles in multilayer hexagonal boron nitride. *2D Materials* **2021**, *8*, 024001.

[19] Zhao, C.;Shan, L.;Sun, R.;Wang, X.; Ding, F. Wrinkle formation in synthesized graphene and 2D materials. *Materials Today* **2024**, *81*, 104-117.

[20] Ares, P.;Wang, Y. B.;Woods, C. R.;Dougherty, J.;Fumagalli, L.;Guinea, F.;Davidovitch, B.; Novoselov, K. S. Van der Waals interaction affects wrinkle formation in two-dimensional materials. *Proceedings of the National Academy of Sciences* **2021**, *118*, e2025870118.

[21] Ishigami, M.;Chen, J.-H.;Cullen, W. G.;Fuhrer, M. S.; Williams, E. D. Atomic structure of graphene on SiO2. *Nano letters* **2007**, *7*, 1643-1648.

[22] Cerda, E.; Mahadevan, L. Geometry and physics of wrinkling. *Physical review letters* **2003**, *90*, 074302.

[23] Bowden, N.;Brittain, S.;Evans, A. G.;Hutchinson, J. W.; Whitesides, G. M. Spontaneous formation of ordered structures in thin films of metals supported on an elastomeric polymer. *nature* **1998**, *393*, 146-149.

[24] Yang, S.;Chen, Y.; Jiang, C. Strain engineering of two-dimensional materials: Methods, properties, and applications. *InfoMat* **2021**, *3*, 397-420.

[25] Toporski, J.;Dieing, T.; Hollricher, O. *Confocal Raman Microscopy*; Springer, 2018.

[26] Itoh, N.; Hanari, N. Reliable evaluation of the lateral resolution of a confocal Raman microscope by using the tungsten-dot array certified reference material. *Analytical Sciences* **2020**, *36*, 1009-1013.

[27] Tran, T. T.;Bray, K.;Ford, M. J.;Toth, M.; Aharonovich, I. Quantum emission from hexagonal boron nitride monolayers. *Nature nanotechnology* **2016**, *11*, 37-41.

[28] Palacios-Berraquero, C.;Kara, D. M.;Montblanch, A. R.-P.;Barbone, M.;Latawiec, P.;Yoon, D.;Ott, A. K.;Loncar, M.;Ferrari, A. C.; Atatüre, M. Large-scale quantum-emitter arrays in atomically thin semiconductors. *Nature communications* **2017**, *8*, 15093.

[29] Xie, Q.;Alibakhshi, M. A.;Jiao, S.;Xu, Z.;Hempel, M.;Kong, J.;Park, H. G.; Duan, C. Fast water transport in graphene nanofluidic channels. *Nature nanotechnology* **2018**, *13*, 238-245.

[30] Melitz, W.;Shen, J.;Kummel, A. C.; Lee, S. Kelvin probe force microscopy and its application. *Surface science reports* **2011**, *66*, 1-27.

[31] Agrawal, K. V.;Shimizu, S.;Drahushuk, L. W.;Kilcoyne, D.; Strano, M. S. Observation of extreme phase transition temperatures of water confined inside isolated carbon nanotubes. *Nature nanotechnology* **2017**, *12*, 267-273.

[32] Fumagalli, L.;Esfandiar, A.;Fabregas, R.;Hu, S.;Ares, P.;Janardanan, A.;Yang, Q.;Radha, B.;Taniguchi, T.; Watanabe, K. Anomalously low dielectric constant of confined water. *Science* **2018**, *360*, 1339-1342.

[33] Fumagalli, L.;Ferrari, G.;Sampietro, M.; Gomila, G. Dielectric-constant measurement of thin insulating films at low frequency by nanoscale capacitance microscopy. *Applied Physics Letters* **2007**, *91*.

[34] Tas, N. R.;Haneveld, J.;Jansen, H. V.;Elwenspoek, M.; van den Berg, A. Capillary filling speed of water in nanochannels. *Applied Physics Letters* **2004**, *85*, 3274-3276.

[35] Squires, T. M.; Quake, S. R. Microfluidics: Fluid physics at the nanoliter scale. *Reviews of modern physics* **2005**, *77*, 977-1026.

[36] Schlaich, A.;Knapp, E. W.; Netz, R. R. Water dielectric effects in planar confinement. *Physical review letters* **2016**, *117*, 048001.

[37] Zhang, K.; Arroyo, M. Understanding and strain-engineering wrinkle networks in supported graphene through simulations. *Journal of the Mechanics and Physics of Solids* **2014**, *72*, 61-74.
[38] Hasted, J.;Ritson, D.; Collie, C. Dielectric properties of aqueous ionic solutions. Parts I and II. *The journal of chemical physics* **1948**, *16*, 1-21.
[39] Geim, A. K.; Grigorieva, I. V. Van der Waals heterostructures. *Nature* **2013**, *499*, 419-425.
[40] Shin, D. H.;Kim, S. H.;Coshic, K.;Watanabe, K.;Taniguchi, T.;Verbiest, G. J.;Caneva, S.;Aksimentiev, A.;Steeneken, P. G.; Joo, C. Diffusion of DNA on Atomically Flat 2D Material Surfaces. *ACS nano* **2025**, *19*, 21307-21318.
[41] Szalai, A. M.;Ferrari, G.;Richter, L.;Hartmann, J.;Kesici, M.-Z.;Ji, B.;Coshic, K.;Dagleish, M. R.;Jaeger, A.; Aksimentiev, A. Single-molecule dynamic structural biology with vertically arranged DNA on a fluorescence microscope. *Nature Methods* **2025**, *22*, 135-144.
[42] Stewart, J. C.;Fan, Y.;Danial, J. S. H.;Goetz, A.;Prasad, A. S.;Burton, O. J.;Alexander-Webber, J. A.;Lee, S. F.;Skoff, S. M.;Babenko, V.; Hofmann, S. Quantum Emitter Localization in Layer-Engineered Hexagonal Boron Nitride. *ACS Nano* **2021**, *15*, 13591-13603.
[43] Radiom, M.;Chan, W.; Yang, C. Capillary filling with the effect of pneumatic pressure of trapped air. *Microfluidics and nanofluidics* **2010**, *9*, 65-75.
[44] Park, S.;Cho, H.;Kim, J.; Han, K.-H. Lateral degassing method for disposable film-chip microfluidic devices. *Membranes* **2021**, *11*, 316.
[45] Choi, S.;Tran, T. T.;Elbadawi, C.;Lobo, C.;Wang, X.;Juodkazis, S.;Seniutinas, G.;Toth, M.; Aharonovich, I. Engineering and Localization of Quantum Emitters in Large Hexagonal Boron Nitride Layers. *ACS Appl Mater Interfaces* **2016**, *8*, 29642-29648.
[46] Jacobs, H.;Knapp, H.;Müller, S.; Stemmer, A. Surface potential mapping: A qualitative material contrast in SPM. *Ultramicroscopy* **1997**, *69*, 39-49.
[47] Collins, L.;Jesse, S.;Kilpatrick, J. I.;Tselev, A.;Varenyk, O.;Okatan, M. B.;Weber, S. A.;Kumar, A.;Balke, N.; Kalinin, S. V. Probing charge screening dynamics and electrochemical processes at the solid–liquid interface with electrochemical force microscopy. *Nature communications* **2014**, *5*, 3871.

**Graphical TOC Entry**

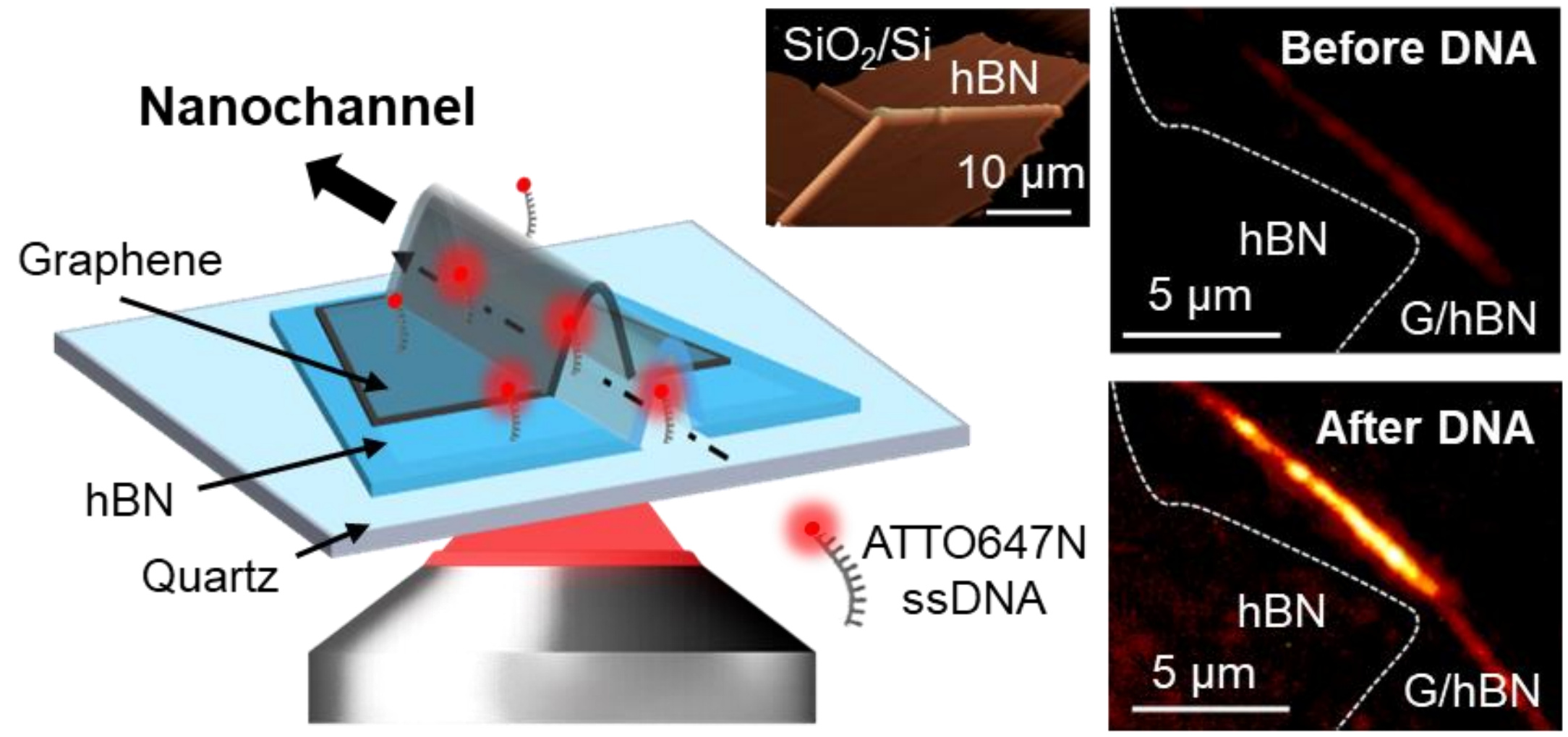

**Wrinkle-induced nanochannels in hexagonal boron nitride act as nanoscale confinements for optofluidics experiments.**

# Supporting Information

# Self-Assembled hBN Wrinkles as Planar Optofluidic Channels

*Xiliang Yang, Tetsuo Martynowicz, Allard Katan, Kenji Watanabe, Takashi Taniguchi, Sabina Caneva*

The supplementary information includes:

## S1. Thermal Expansion Coefficients of Substrates and 2D Materials

Table S1 lists the in-plane thermal expansion coefficients (TEC) of common substrates (sapphire, $SiO_2$/Si, quartz) and 2D materials (hBN, graphene). Positive TECs indicate thermal expansion, while negative TEC indicates in-plane contraction with increasing temperature. The TEC mismatch between positive-expanding substrates and negative-expanding 2D materials induces compressive strain, driving wrinkle formation during thermal processing.

**Table S1.** In-plane thermal expansion coefficients (TEC) of typical substrates and 2D materials

| Sapphire | $SiO_2$/Si | Quartz | ITO | hBN | Graphene |
|---|---|---|---|---|---|
| +7.5–8.5 | +1-2 | +0.55 | ~7.2$\pm$0.3 | –2 to –1 | –8 to –1 |

Note: Values are in $\times 10^{-6}$ $K^{-1}$, typically measured between 300–1000 K.

## S2. Raman–PL correlation at wrinkle junctions: strain and wavelength-dependent emission.

After high-temperature annealing, multilayer hBN forms wrinkle networks with frequent three-way junctions. To link strain localization with emitter wavelength, we combined spatially resolved Raman mapping with wavelength-selected PL mapping at a representative junction (Fig. S2). The hBN E2g peak shows systematic spatial shifts around the junction, indicating an inhomogeneous strain landscape associated with the wrinkle geometry. Consistent with this, the PL maps reveal wavelength-dependent localization: emission around 565 nm is primarily concentrated on the wrinkle ridges/crest regions, whereas emission around 620 nm is preferentially found along the wrinkle edge regions. These results support that wrinkle-induced strain fields correlate with spectral tuning and spatial selection of emitters in hBN, and that junction geometries can serve as strain-defined nodes for engineering multi-wavelength emission within a connected wrinkle network.

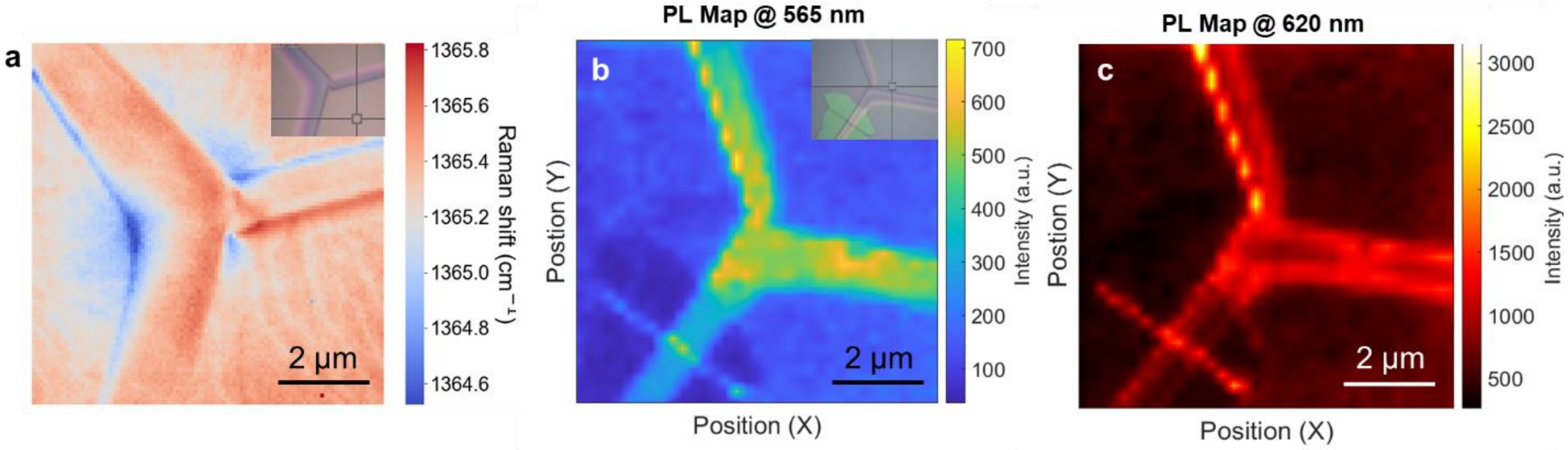

**Figure S2. Strain localization and wavelength-dependent emitter distribution at a three-way wrinkle junction. (a)** Raman map of the hBN $E_{2g}$ peak position showing localized strain variations around the junction (flake 1). **(b,c)** Wavelength-selected PL intensity maps at 565 nm and 620 nm, respectively, highlighting distinct spatial localization of emitters with different peak wavelengths (flake 2).

### S3. Dielectric imaging of wrinkle-confined liquids: air, water, and buffer.

Topographic and capacitance-gradient (dC/dz) AFM images were sequentially acquired from the same wrinkle region in air, Milli-Q water, and 1× TAE buffer containing 10 mM $MgCl_2$. Upon water filling ($\varepsilon \approx 80$), the dC/dz contrast became strongly negative (relative to the surrounding terrace background), consistent with enhanced capacitive coupling through the high-dielectric medium. After replacing water with buffer ($\varepsilon \approx 70$), the contrast magnitude decreased, reflecting the lower permittivity of the confined liquid. Although minor changes in tip condition during liquid exchange cannot be completely excluded, the systematic and reversible evolution of the signal with dielectric constant supports that the observed contrast arises primarily from variations in local permittivity within the confined liquid rather than from surface contamination or topographic artifacts.

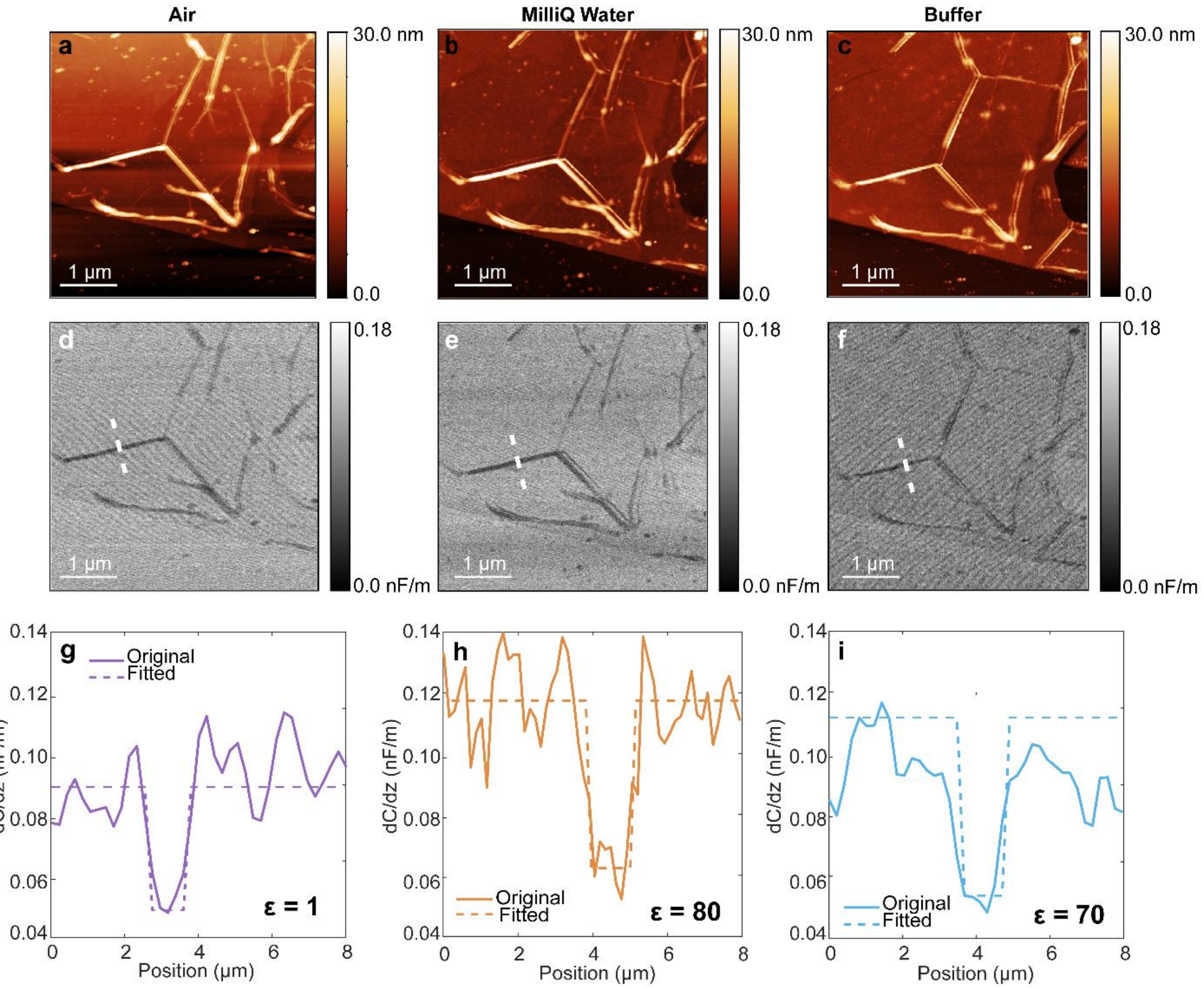

**Figure S3. Dielectric imaging of wrinkle-confined liquids: air, water, and TAE/MgCl₂ buffer.** **(a–c)** Topographic AFM images of the same wrinkles in air, filled with water, and filled with 1× TAE buffer containing 10 mM $MgCl_2$, respectively. **(d–f)** Capacitance gradient (dC/dz) images of the wrinkles under the same conditions. **(g-i)** dC/dz line profiles across the wrinkles (white dashed lines in b-f); solid lines are measured data, dashed lines are fitted curves.

## S4. Fluorescence and AFM characterization of ssDNA in hBN wrinkles.

Fluorescence and topographic AFM images of hBN wrinkles after deposition of fluorescently labeled single-stranded DNA (ssDNA). Bright emission spots often coincide with wrinkle regions but are also observed along the flat surface, indicating that part of the signal originates from DNA adsorbed on top of the wrinkles rather than solely from molecules confined within them. These results suggest that the observed fluorescence cannot be unambiguously attributed to DNA inside the wrinkle channels, as both surface binding and partial confinement contribute to the overall emission intensity. Accordingly, this dataset is used as qualitative evidence of wrinkle-associated localization rather than definitive proof of in-channel confinement.

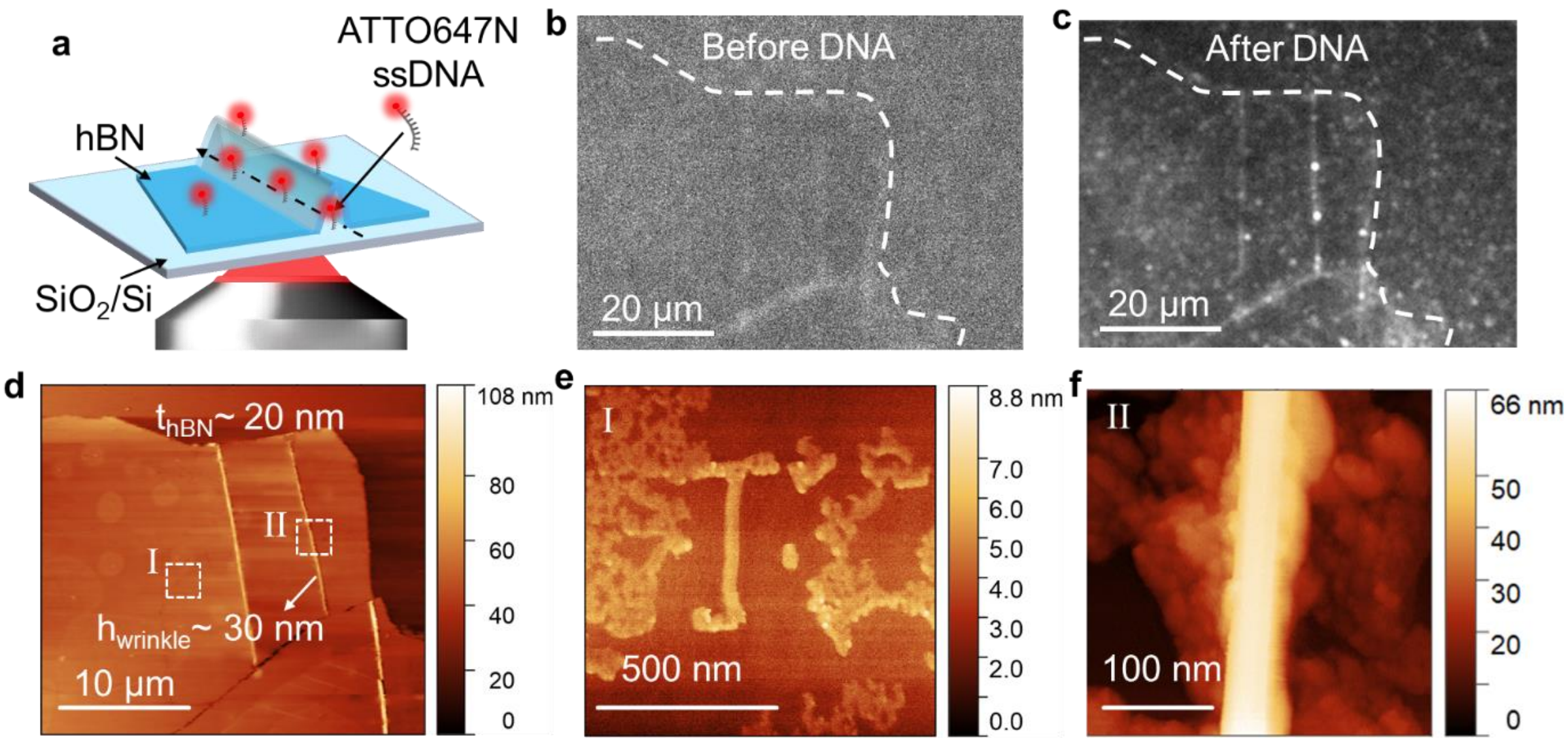

**Figure S4. Fluorescence and AFM imaging of ssDNA in wrinkle-confined hBN channels. (a)** Schematic illustration of ATTO647-labeled ssDNA interacting with a wrinkle-induced hBN nanochannel, showing possible localization both inside the wrinkle and on the surface. **(b)** Epi-fluorescence image of the wrinkles before DNA deposition. **(c)** Epi-fluorescence image after DNA deposition, showing stable fluorescent spots at wrinkle locations and in flat areas. **(d)** AFM topography of the same region after DNA deposition. **(e)** Zoom-in of area 1 in (d), showing clusters of ssDNA molecules adsorbed on the pristine hBN surface. **(f)** Zoom-in of area 2 in (d), showing material located on top of wrinkles.

## S5. Comparison of blocked and open wrinkle pathways for ssDNA loading

We compare two representative wrinkle segments within hBN/graphene overlap regions: one exhibiting a locally blocked/non-through segment and one exhibiting an open, through-connected pathway. The comparison highlights that continuous fluorescence localization requires an uninterrupted capillary route into the wrinkle-defined cavity.

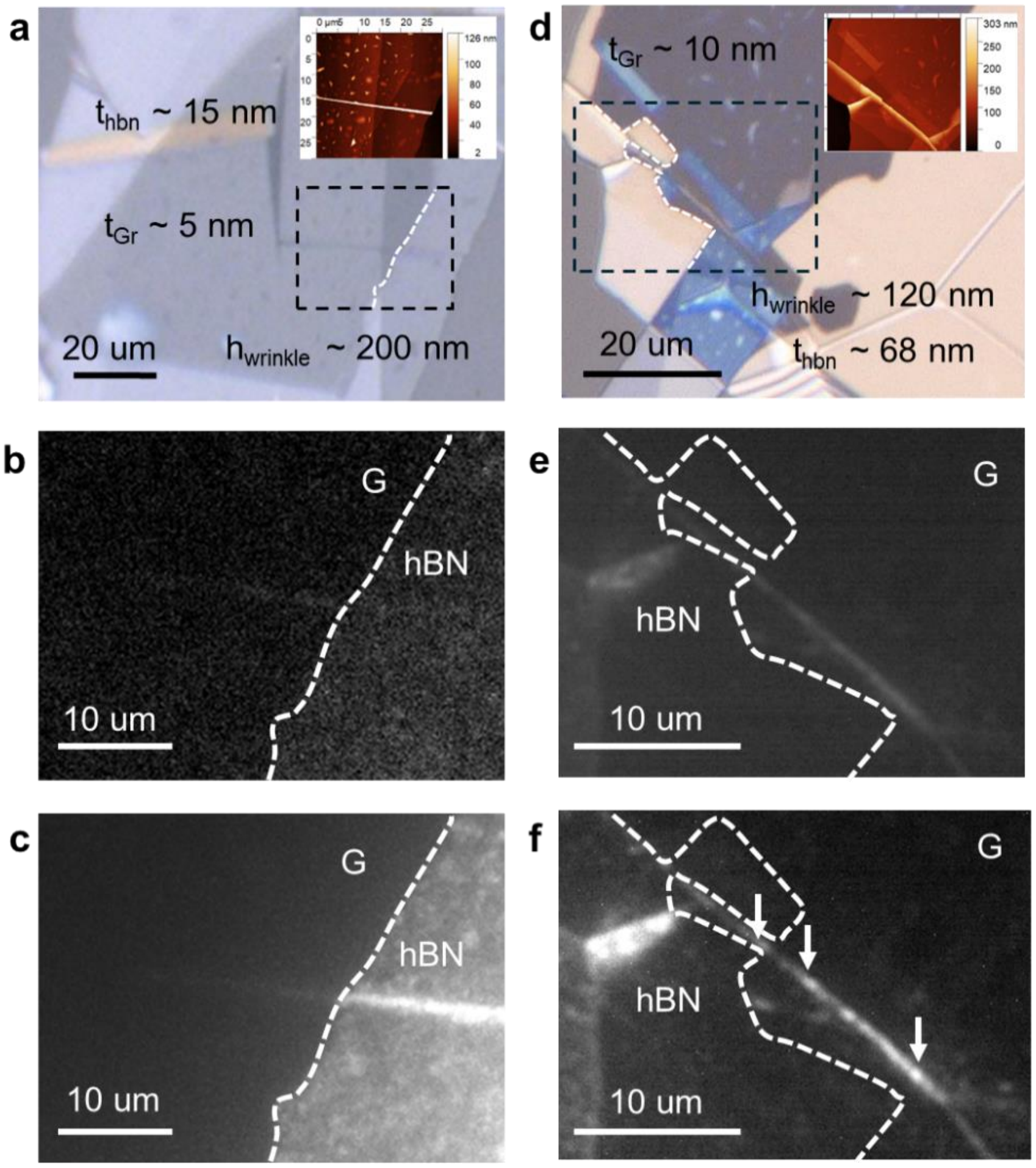

**Figure S5. Comparison of locally blocked and open wrinkle pathways for ssDNA loading in hBN/graphene stacks. (a,d)** Bright-field optical images of two representative hBN/graphene overlap regions. Dashed boxes indicate the fluorescence imaging areas shown in (b,e). Insets: AFM topography of the corresponding wrinkle segments; labels indicate the hBN and graphene

thicknesses and a representative wrinkle height. In (a), a blocked region is marked to denote a locally non-through/obstructed segment along the nominal route. **(b,e)** Epifluorescence images of the same regions before ssDNA addition (background reference). Dashed outlines mark the hBN/graphene boundary. **(c,f)** Epifluorescence images acquired after loading and drying ATTO647N–ssDNA. (a–c) In the locally blocked case, no continuous fluorescence develops along the expected pathway. (d–f) In the open, through-connected case, fluorescence localizes continuously along the wrinkle-defined route (arrows), consistent with capillary access and retention within the wrinkle cavity.

**S6. Fluorescence of hBN wrinkles before and after buffer immersion.**

Control experiments confirmed that buffer alone produces no detectable fluorescence and does not activate or modify intrinsic hBN emission, demonstrating the optical stability of wrinkle-induced channels. The sharpening of the features in (c) with respect to panel (b) is consistent with improved refractive-index matching when imaging through buffer with an oil-immersion objective compared to air.

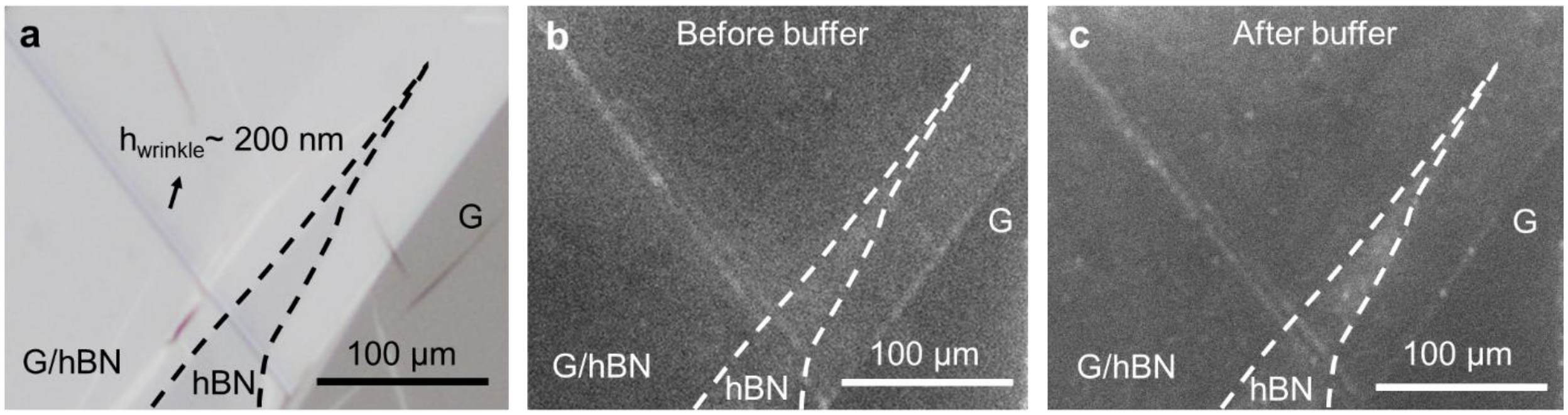

**Figure S6. Optical stability of hBN wrinkles under buffer immersion. (a)** Optical image of a graphene/hBN stack (hBN thickness ∼70 nm, graphene ∼12 nm). **(b)** Epi-fluorescence image of a wrinkle under 635 nm excitation before buffer immersion. **(c)** Epi-fluorescence image of the same wrinkle after buffer diffusion, showing unchanged emission after drying.